\documentstyle[aps,twocolumn,psfig]{revtex}
\begin{document}
\draft
\preprint{}
\title{
Anomalous Behavior of Ru for Catalytic Oxidation:\\
A Theoretical Study of the Catalytic Reaction
${\rm CO} + \frac{1}{2} {\rm O}_2 \longrightarrow {\rm CO}_2$
 }
\author{C. Stampfl and M. Scheffler}
\address{
Fritz-Haber-Institut der Max-Planck-Gesellschaft,
Faradayweg 4-6, D-14195 Berlin-Dahlem, Germany
}
\date{Received 15 April 1996}

\twocolumn[

\maketitle

\vspace*{-10pt}
\vspace*{-0.7cm}
\begin{quote}
\parbox{16cm}{\small
 Recent experiments revealed an  anomalous dependence of
 carbon monoxide oxidation at Ru\,(0001) on oxygen pressure and 
 a particularly high reaction rate.
Below we report density functional theory calculations of the
energetics and reaction pathways of the speculated mechanism.
We will show that the exceptionally high rate is actuated by
a weakly but nevertheless well bound $(1 \times 1)$ oxygen adsorbate layer.
Furthermore it is 
found  that reactions via  scattering of 
{\em gas-phase}
CO at the oxygen covered surface may play an important role.
Our analysis reveals, however, that reactions 
via {\em adsorbed} CO molecules (the so-called Langmuir-Hinshelwood mechanism)
dominate.\\

} 
\end{quote}
]
 
\narrowtext
The oxidation of carbon monoxide at transition metal surfaces
is one of the most extensively studied heterogeneous catalytic
reactions (see for example
   \cite{king,oxygen,peden1}
and references therein).
Numerous investigations performed under ultra high vacuum (UHV) conditions
for many different metal surfaces have shown that the reaction
proceeds  via the so-called Langmuir-Hinshelwood (L-H) mechanism, which
means that the reaction takes place between {\em chemisorbed} reagents. 
Typical conditions for working catalysts, however, are  higher
pressure and temperature. Although it has been demonstrated that for
a number of systems extrapolation of data over a wide pressure
range is valid~\cite{topsoe}, such a conclusion cannot be generalized.
Recent high gas pressure studies (e.g.
at about 10 torr),
for oxidizing conditions
(i.e., at CO/O$_{2}$ pressure ratios $ < 1$)
\cite{peden1,peden2,peden3}
reported that CO$_{2}$ production over Ru\,(0001) is anomalous: 
\\[-0.5cm]
\begin{itemize}
   \item [1)] The rate of CO$_{2}$ production
was found to be exceptionally high -- significantly higher than at any other
transition metal surface. Interestingly, under UHV Ru\,(0001) is by far the
poorest catalyst~\cite{king}. 
\\[-0.5cm]
 \item [2)] The measured kinetic data (activation energy and
pressure dependencies) were found to be markedly different to those for
other substrates, and in contrast to the other transition metal catalysts,
Pt, Pd, Ir, and Rh, 
 highest rates occurred for high concentrations of oxygen at
the surface.
\\[-0.5cm]
    \item [3)] Almost no chemisorbed CO was detected during or after
the reaction.
\\[-0.5cm]
\end{itemize}
It was therefore speculated that 
the Eley-Rideal (E-R) mechanism is operational as opposed to the
``usual'' L-H mechanism.
In the E-R mechanism, the reaction occurs  between
{\em gas-phase} and chemisorbed particles.
So far E-R mechanisms have only been observed experimentally
for somewhat artificial reactions triggered by a  beam
of atomic hydrogen (or deuterium)~\cite{rettner}. 
To gain
understanding into the drastically different behavior of Ru\,(0001)
for the CO oxidation reaction,  we carried out density functional
theory (DFT) calculations, where the main approximations are
the
supercell approach  and the employed functional for the
exchange-correlation interaction.
For the latter we use the generalized gradient 
approximation (GGA)~\cite{perdew} which is 
the best justified treatment to date.
Our study represents the first theoretical attempt to follow a
heterogenous catalytic reaction (molecular and dissociative (atomic) 
adsorption, surface reaction,
desorption of products)  using DFT-GGA and an extended surface. 
In brief, the details of the theoretical approach are summarized as follows:
We use {\em ab initio}, fully separable, norm-conserving
DFT-GGA pseudopotentials~\cite{troullier}.
The GGA is thus treated in a consistent way, from the free atom to the
solid surface and the reactants.
Relativistic effects are taken into account
by using spin averaged potentials. 
The surface calculations are
performed using a $(2 \times 2)$ surface unit cell, a four atomic
layer Ru slab, and a vacuum region corresponding to thirteen such layers.
We use an energy cut-off  of 40 Ry with three special
{\bf k}-points~\cite{cunningham}
in the two-dimensional Brillouin zone.
Convergence tests for O on Ru\,(0001)
indicated that this basis set provides a sufficiently accurate 
description ~\cite{stampfl}.
The adsorbate structures are created on one side of the slab~\cite{neugebauer}
where we relax the position of all the atoms using a damped
molecular dynamics~\cite{stumpf}, except for
the Ru atoms in the bottom two layers, which are kept at their bulk-like
positions.
Details of the calculations will be published elsewhere~\cite{paper}.

 It is well known that
under UHV conditions,
at room temperature, dissociative adsorption of O$_2$
results in a saturation coverage of $\Theta_{\rm O} \approx 1/2$
corresponding to the formation of a $(2 \times 1)$ structure~\cite{pfnur}.
Recently, from DFT-GGA calculations we predicted that
an even higher coverage should be stable on the surface,
namely,  a ($1 \times 1$) structure with coverage  $\Theta_{\rm O} = 1$,
where the O atoms occupy hcp-hollow sites~\cite{stampfl}.
Subsequently, this structure was indeed successfully created
under UHV conditions~\cite{over}, where
it was concluded that formation of the $\Theta_{\rm O} = 1$ structure
from gas-phase O$_2$ is hindered kinetically~\cite{stampfl,over}, but
by offering atomic oxygen (or under high pressure conditions),
this phase can be attained. 
We also carried out
calculations involving higher oxygen coverages on the
surface, as well as geometries involving sub-surface
oxygen; these structures were
found to be unstable and metastable, respectively,
with respect to gas-phase O$_{2}$. We therefore do not expect them to
play an important role for the present investigation.
We refer to Refs.~\cite{over,stampfl2} for more details.
Because the conditions under which the particularly high rates of CO$_{2}$
formation occur involve elevated partial gas pressures (and CO/O$_{2}$ 
ratios $<$ 1), there will be a  significant attempt frequency
to overcome activation barriers for dissociative adsorption of O$_{2}$.
Thus, it is likely that during reaction 
the oxygen coverage on the surface approaches one 
monolayer.
We therefore initially assume in our investigation of the
oxidation of CO at Ru\,(0001)
that the $(1 \times1 )$ phase covers the surface.

As mentioned above, from the experiments it had been speculated that
CO may react from the gas-phase with adsorbed oxygen 
(the E-R mechanism).
To investigate this
possibility we first ask whether CO  can adsorb on the
$(1 \times 1)$-O/Ru\,(0001) surface.
The sites considered were the on-top and  fcc-hollow sites,
with respect to the Ru\,(0001) substrate, and a bridge site between two adsorbed
O atoms (compare inset of Fig.~~1).
For each site we calculated the energy as a function
of distance of the molecule from the surface.
In these calculations the CO-axis is held
perpendicular to the surface with the C-end of the molecule
closest to the surface.
At each point we fix the position of the C atom and relax
the positions of all the O atoms and the top two Ru layers.
The results are shown in the left panel of Fig.~1, where we have 
also considered the path for CO directly above an adsorbed O atom.
It can be seen that CO experiences an energy barrier
which starts to build up
at about 2.5~\AA\,
from the surface for all sites, reflecting
a {\em repulsive} interaction with the O-covered surface.
Furthermore, it is apparent that the surface potential is 
significantly  corrugated: considering
a constant-total-energy surface
as a function of the lateral position of the CO, we find that it 
exhibits the lowest energy (but always repulsive) over
the fcc-hollow site. Thus CO tends to avoid the O adatoms
but will not form a chemical bond with the metal substrate.
The O-covered surface thus {\em prevents}
reaction via the L-H process leaving the possibility for 
reaction via {\em gas-phase} CO with chemisorbed O, i.e. the E-R reaction.
In this respect,
for the approach of CO directly above an adsorbed O atom (full circles
in Fig.~1),
we find that beyond a critical distance, the repulsive interaction
turns into an attractive one,
and the CO molecule and the adsorbed oxygen atom
react to form CO$_{2}$.
On relaxing the position of the C-atom,
the CO$_{2}$ molecule then leaves the surface
with a significant  energy gain of $\approx$1.95~eV.
The associated energetics are shown in the right panel of Fig.~1.
It is important here to emphasise that we have considered all relevant reaction
paths for CO at the $(1 \times 1)$-O/Ru\,(0001) surface and that although
on first consideration, Fig.~1 may appear to 
suggest that the favored
reaction pathway for CO$_{2}$ formation is 
over sites away from the adsorbed O atom, in particular the
fcc-hollow site, this is not the case:
Indeed ``slow moving'' CO molecules with low translational energy will be
``steered'' towards the fcc-hollow sites. These molecules will however not
achieve reaction due to the sizeable energy barrier.
Fast CO molecules of high translational energy, not susceptable to steering
effects, which are incident at sites away from an adsorbed O atom will also
not react, but will rather be reflected from 
the surface.
Thus, to produce CO$_{2}$ via this mechanism, the results indicate 
that the molecule must ``hit'', or get very close to, an adsorbed O atom.
Interestingly, the calculations show that there is a physisorption
well for CO, as well as for CO$_{2}$, above the surface (barely visible
in Fig.~1).
The wells are very shallow ($\approx 0.04$~eV) and thus
they will not play a role. It should be  noted, however,
that the calculated depths are likely to be  lower bounds because the employed
exchange-correlation functional does not describe the long-range
(van der Waals type) interactions and
the physisorption wells are found at distances
where the true potential energy is likely to be
more attractive than that given by
the DFT-GGA calculation. 

A more detailed understanding of the pathway for reaction via scattering of
CO is obtained by
evaluating an appropriate cut through the high-dimensional
potential energy surface (PES);
this cut is defined by two variables: the
vertical position of the C atom
and the vertical position of the O adatom below the molecule.
In order of ease of analysis,
the CO-axis is initially held perpendicular to the surface.
The resulting PES  is presented in Fig.~2, where
the coordinate system is shown as the inset.
For each point we relaxed all the O atoms (except 
that held fixed at $Z_{\rm O}$), and the top
two Ru layers.
The repulsive interaction is again evident as CO nears the surface.
In response to the approaching molecule, the O adatom 
moves in towards the surface:
For example, at a distance of $Z_{\rm C} = 1.9$~\AA\,, the O atom is
displaced inwards by $Z_{\rm O} = 0.2$~\AA\,.
Thus, the impinging CO molecule
``hits'' a ``soft wall''.
Reaction to CO$_{2}$ is achieved via an upward movement
or ``hop'' of the O adatom by $\approx 0.4$ \AA\, towards the CO molecule
(corresponding to  movement parallel to the horizontal axis of Fig.~2) and
brings the system to the transition state of the reaction marked by the
asterisk.
In view of the similar masses of O and C, it is likely that the
impinging CO molecule will impart a significant amount of energy to the O
adatom, thus stimulating its vibrations and facilitating its motion
(indicated by the oscillations in the dot-dashed curve).
The newly formed CO$_{2}$ molecule then finds itself in a
particularly unfavorable position and is strongly repelled
from the surface towards the vacuum region
with a large energy gain
of 1.95~eV.
In the cut through the PES shown in Fig.~2, the energy barrier
hindering CO$_2$ formation is  $\approx$1.6~eV.

The PES of Fig.~2 corresponds to a constrained
situation of the surface--CO angle. When this constraint is dropped, i.e.,
when the tilt angle of the CO-axis
is allowed to relax~\cite{comment1}, we find that
the energy barrier is reduced to 1.1~eV, and
also that the position of the saddle point of the PES occurs closer  to the
surface (by 0.3~\AA).
At the transition state (see Fig.~3), the optimum tilt angle with respect to the
surface normal is found to be $49^{\rm o}$  which
corresponds to a bond angle
of  $131^{\circ}$ for the ``CO$_{2}$-like'' complex.
Interestingly, this geometry is very similar to that associated
with the CO$_{2}^{-}$ ion \cite{bagus} and to that proposed
for the ``activated complex'' for the CO oxidation reaction over
other transition metal surfaces \cite{coulston}.

We have thus identified a likely reaction pathway for the
E-R mechanism. The activation energy barrier for this type of 
reaction appears to be sizable. However, it is very similar to those 
derived from experimental studies of CO oxidation reactions at other
surfaces~\cite{peden1}
which proceed via a L-H process, and also for the
measurements at Ru\,(0001) the estimated activation energy
is comparable, namely 0.85 eV~\cite{peden1,peden2,peden3}.
On the basis of the present results we predict an energy diagram
for the E-R mechanism, which  is shown in Fig.~4.
An estimate of the reaction rate gives 
$R = 7.5 \times 10^{6} \exp(-1.1/(k_{\rm  B}T)) \,s^{-1}$ which yields
at $T=500$~K, $R= 6\times 10^{-5}$
CO$_{2}$ molecules formed per surface Ru atom per second
\cite{comment2} which is about 
$3 \times 10^{-6}$ smaller than that observed experimentally~\cite{peden2}. 
This indicates that this mechanism
alone cannot explain the particularly high CO$_{2}$
turnover rate.
Nevertheless, the rate is only about a factor of 10$^{-3}$ 
less than that for the L-H process at Pt or Pd \cite{peden1} and
with {\em molecular beam} experiments
this predicted E-R mechanism and associated energetics could possibly be 
measured for the first time for the CO oxidation reaction.

To understand the high reaction rate reported  experimentally, we turn to
another consideration:
CO molecules might adsorb at sites at which
an oxygen atom has been removed (e.g. by the above described E-R reaction).
Indeed, assuming thermal equilibrium of the
CO + O$_2$ gas and a mixed CO + O adlayer, the law of mass
action indicates 
that about 0.03~\% of the sites of the $(1 \times 1)$ adlayer will be
occupied by CO (we assumed that the
O$_2$ and CO partial pressures are equal, the temperature
is $T = 500$~K, 
and the binding energy of CO into an O-vacancy of the
$(1 \times 1)$ adlayer is calculated to be 0.85~eV 
and the adsorption
energy of $\frac{1}{2}$ O$_2$ (i.e. an O atom) into a vacancy is
1.20~eV.
In reality the CO concentration will be even higher
because catalysis is not ruled by thermodynamic equilibrium but by kinetics,
and we find that CO adsorption into an existing O vacancy
can proceed basically without hindrance while O adsorption (from O$_2$)
is hindered by an energy barrier (see, e.g. Ref.~\cite{over}).
Therefore the actual percentage of surface sites occupied by CO will 
be somewhat larger.
For these CO molecules there is a substantial attempt
frequency to form a CO$_2$ molecule with neighboring O adatoms,
now by the L-H mechanism which we expect to proceed very efficiently
due to the relatively weak
binding energy of {\em both} CO (which we calculate to be 
about half that which it has on the clean
surface, -- and on the surface with O-coverages: $\Theta \leq$ 0.5) and
O atoms in the high-coverage $(1 \times 1)$ adlayer, as well as
the close proximity of the constituents.
We find that the energy gain on CO$_{2}$ formation
(of the surface reaction), via this mechanism
is about 0.66~eV~\cite{comment3}; 
noticeably smaller than that of 1.95~eV (see Fig.~4)
but still quite significant if compared to that of
$\approx 0.2$~eV at Pt\,(111) and Pd\,(111) \cite{king} as determined
experimentally.

In summary, we now have the following picture of CO oxidation at Ru\,(0001):
with respect to other transition metals, ruthenium binds oxygen 
 particularly strongly.
Therefore, at low oxygen coverages a Ru catalyst dissociates O$_2$
efficiently, but (in contrast to e.g. Pd) it holds the oxygen (and CO)
so strongly that  reaction to CO$_2$
is disfavored.
A good catalyst should actuate this
dissociation but at the same time should not bind the dissociated 
entities too strongly which gives them good capability to diffuse and react. 
Too strongly bound constituents would have little reason to react at all.
For oxygen 
in the $(1 \times 1)$-O monolayer, the adsorption energy
is significantly weaker and thus CO$_2$ formation enhanced.
Our results  indicate that this
high coverage oxygen phase enables reaction via
both scattering of CO
(the E-R mechanism) and by the L-H mechanism where
the former 
may play an important role in initiating the reaction.
The high rate then  develops and is maintained by an efficient L-H mechanism.
Our theoretical results thus explain the anomalous dependence of the
reaction on oxygen pressure, as only under sufficiently high oxygen pressure
the $(1 \times 1)$ layer is attained. 

The different mechanisms identified in our study are also likely
to play important roles for other catalytic reactions.
We hope
that the detailed predictions and the
unusual mechanism outlined above, will be tested by additional
experiments.

We wish to thank Martin Fuchs for his
help in creating the pseudopotentials.

\newpage

\begin{figure}
\psfig{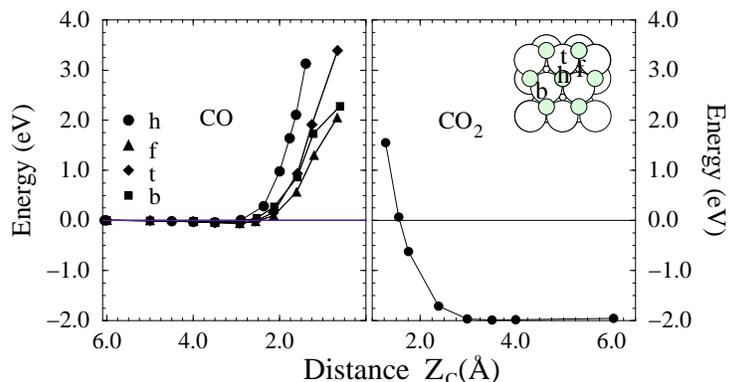}	
\caption{Energy as a function of distance of the C atom, $Z_{\rm C}$, 
of the  CO and CO$_{2}$ molecules
from the surface for the various sites tested.
The molecular axes are constrained to be perpendicular to the surface.
The zero of energy refers to the
situation where CO is far  away from the $(1 \times 1)$-O/Ru\,(0001)
 surface ($Z_{\rm C}\protect\approx$6~\AA\,).}
\end{figure}

\begin{figure}
\psfig{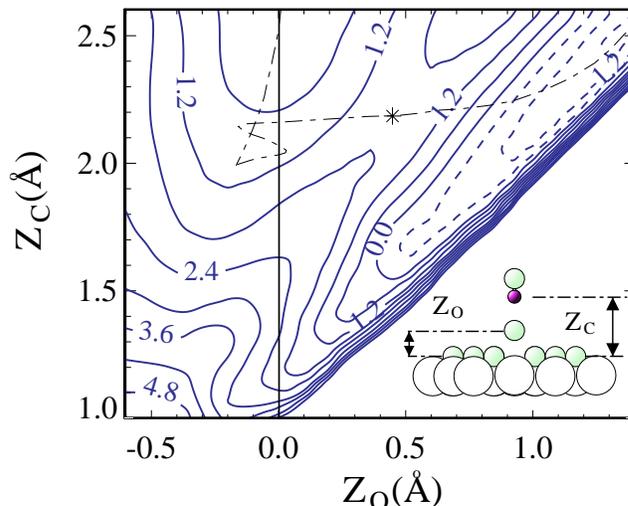}		
\caption{Cut through the high-dimensional potential energy surface (PES) 
as a function of the positions of the C atom, $Z_{\rm C}$, 
and the O adatom, $Z_{\rm O}$ (see inset).
The molecular axes are constrained to be perpendicular to the surface.
Positive energies are shown as continuous lines, negative ones as
dashed lines.
The contour-line spacing is 0.6 eV. 
The dot-dashed line indicates a possible reaction pathway.}
\end{figure}
\clearpage
\begin{figure}
\psfig{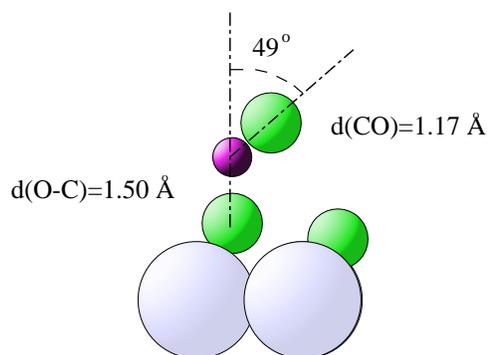}
\caption{Transition state geometry identified
for the  reaction of gas-phase CO
with adsorbed oxygen when the constraint on the molecular axis
is relaxed. The large, medium, and small circles represent Ru, O, and
C atoms, respectively.}
\end{figure}

\begin{figure}
\psfig{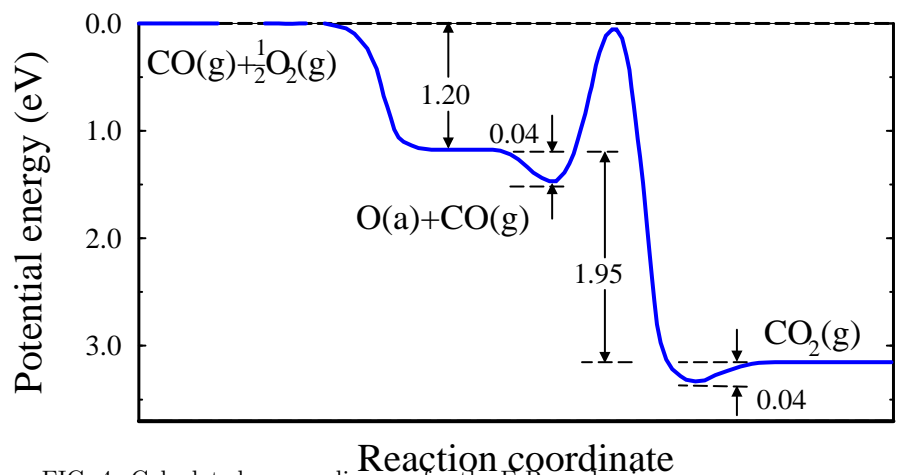}	
\caption{Calculated energy diagram for the E-R mechanism of CO oxidation
at Ru\,(0001). Note that the depths of the physisorption wells are
exaggerated for clarity. }
\end{figure}


\begin{references}
\bibitem{comment1}
Tilts of the CO-axis were optimized in two planes, with
essentially the same result. The planes were those
passing through the fcc-hollow and the on-top sites. The lateral position of
the C atom was fixed to be the same as that of
the O adatom.
\bibitem{comment2} For
our estimate we use the frequency of impinging CO molecules  per surface
Ru atom of
$\nu = 7.5 \times 10^{6}\, s^{-1}$
which corresponds to a CO gas phase pressure of 16 torr.
\bibitem{comment3} This value is obtained as the reaction energy of
CO + $\frac{1}{2}$O$_{2} \longrightarrow $CO$_{2}$ in vacuum, minus 
the energy gained
on adsorption of $\frac{1}{2}$O$_{2}$ and CO onto the surface.
\end{references}
\end{document}